\documentstyle[sprocl]{article}

\newcommand{\be}{\begin{eqnarray}}
\newcommand{\ee}{\end{eqnarray}}
\newcommand{\p}{\partial}

\newcommand{\cD}{{\cal D}}

\newcommand{\bi}{\bigskip}
\newcommand{\no}{\noindent}
\newcommand{\Zset}{{\sf Z}\hskip -5.5pt{\sf Z}}
\newcommand{\rk}{\right)}
\newcommand{\lk}{\left(}

\begin{document}

\title {Dual description of QCD\footnote{Talk presented at the conference
``Quark Confinement and the Hadron Spectrum II'' at Villa Olmo,
Como, Italy, June 26-29, 1996}}

\author{H. Reinhardt}

\address{Institut f\"ur Theoretische Physik, Universit\"at 
T\"ubingen,\\ Auf der Morgenstelle 14, D-72076 T\"ubingen, Germany}

\maketitle

\abstracts{
It is demonstrated that the field strength approach to Yang
Mills theories has essential features of the dual description. In
D=3 this approach is formulated in terms of gauge invariant
variables.}

\section{Introduction}

According to 't Hooft and Mandelstam~\cite{r1}
confinement may be realized as dual Meissner effect. This
confinement scenario assumes that the QCD ground state consists
of a condensate of magnetic monopoles (dual superconductor),
which squeezes the color electric field of color charges into
flux tubes. This scenario has indeed received support from
lattice calculations~\cite{r2}.

Obviously the dual Meissner effect can be most efficiently
described in a dual formulation, which is known to exist for quantum
electrodynamics. The transition to the dual theory basically
amounts to an interchange of the electric and the magnetic
fields. At the same time the coupling constant is inverted.
In non-Abelian gauge theories duality was considered first by
Montone and Olive~\cite{r3} who conjectured that solitons of the original
gauge theory become massive fields in the dual theory. This idea
was taken up by Seiberg and Witten~\cite{r4} who studied duality in
supersymmetric theories. By showing that certain supersymmetric
models are dual to each other they succeeded to find exact
solutions to the strong coupling regime of some
supersymmetric gauge theories.

Obviously the strong coupling regime of QCD could be most
efficiently studied within the dual theory. Unfortunately the
dual theory of non-super\-sym\-metric Yang-Mills theory and in
particular of QCD is not known and perhaps does not exist in the
strict sense. For this reason there have been attempts to
construct the dual theory of QCD phenomenologically~\cite{r5}. 
There is also a microscopic approach to this problem which has
not been fully appreciated in the past. This is the so-called
field strength approach~\cite{r6} which formulates the
Yang-Mills theory in terms of field strengths. In my talk I
would like to demonstrate that this approach in fact yields a
formulation of Yang-Mills theory which comes very close to a
dual description.

\section{The field strength approach as dual formulation
of Yang-Mills theory}

Consider the standard functional integral formulation of Yang-Mills
theory 
\be
\label{g1}
Z[j] = \int DA_\mu \delta_{gf}\exp\left[-{1\over 4 \kappa^2}
    \int (F(A))^2 + \int jA\right]~.
\ee
Here  $A_\mu(x)$ denotes the gauge field, $j$ is an external
source and $\delta_{gf}$ is a short notation for the gauge
fixing. Furthermore
$F_{\mu\nu}^a(A) = \p_\mu A_\nu^a - \p_\nu A_\mu^a
+ f^{abc} A_{\mu}^b A_\nu^c$
is the field strength where $f^{abc}$ denotes the structure
constant of the gauge group SU(N) and $\kappa$ is the coupling
constant. Let us emphasize that due to the gauge fixing the
measure of the functional integral is not flat. In fact recently
it was explicitly demonstrated that the standard Faddeev-Propov
gauge fixing yields precisely the required Haar-measure for the
gauge invariant partition function~\cite{r7} (see also
ref. 8).

In the field strength approach the Yang-Mills action is
linearized by means of an auxiliary tensor field
$\chi_{\mu\nu}^a(x)$, which has the structure of a field
strength, by the identity~\cite{r6}
\be
\label{g3}
\exp \left[ -{1\over 4 \kappa^2} \int \lk F(A)\rk^2\right]
= \int D \chi_{\mu\nu}^a \exp \left[ -{\kappa^2\over 4}
\int \chi^2 + {i\over 2} \int \chi F(A)\right]~.
\ee
Inserting this identity into the functional integral ({\ref{g1})
and casting the gauge fixing constraint from the gauge potential 
to the tensor field $\chi_{\mu\nu}^a(x)$ one can integrate out 
the gauge field explicitly leaving an effective tensor theory 
defined by
\be
\label{g4}
Z[j] = \int D\chi_{\mu\nu} \lk \det \hat\chi\rk^{-1/2}
\exp \left[ -S_{FS}(\chi)-S_j(\chi)\right] ~.
\ee
Here the functional determinant of the matrix
$\hat\chi_{\mu\nu}^{ab} = f^{abc} \chi_{\mu\nu}^c$ arises from
the Gaussian integral over the gauge field. The action of the tensor
field 
\be
\label{g5}
S_{FS}(\chi) = {\kappa^2\over 4} \int \chi^2 + {i\over 2}
\int \chi F(V)
\ee
is just the exponent of the right-hand side of equation
(\ref{g3}) taken, however, at the stationary phase value of the
gauge potential, which is given by the induced gauge potential 
\be
\label{g6}
V_\mu^a = \lk \hat\chi^{-1}\rk^{ab}_{\mu\nu}
\p_\lambda \chi_{\nu\lambda}^b~.
\ee
This induced gauge potential behaves under gauge transformation
like the original gauge field $A_\mu(x)$. Finally,
\be
\label{g7}
S_j(\chi) = \int jV + {i\over 2} \int j \hat\chi^{-1}j
\ee
contains the dependence on the external source $j$.

Let us emphasize that (\ref{g4}) is an exact representation of
the initial Yang-Mills functional integral equation (\ref{g1}).
A few comments are here in order. From electrodynamics we know
that certain phenomena in topologically non-connected spaces cannot be
described exlusively in terms of the field strength but are
sensitive to the (at least topological) properties of the gauge
potential, as in the case of the Bohm-Aharonov effect. Therefore
one might wonder how the field strength formulation (\ref{g4})
can be an equivalent representation of the Yang-Mills theory
(\ref{g1}). In fact, due to the presence of the induced gauge
potential (\ref{g6}), which in fact couples to the external source
$j$ in the same way as the initial gauge potential, the field
strength approach (\ref{g4}) is also capable of describing those
phenomena. 

Let us compare now the field strength formulation with the
standard formulation. In the standard formulation we start with
the gauge potential $A_\mu(x)$ and construct from this potential
the field strength $F_{\mu\nu}(A)$. By construction this field
strength then satisfies the Bianchi identity 
\be
\label{g8}
   \left[ D_\mu(A), \tilde F_{\mu\nu}(A) \right] =0~,
\ee
where 
\be 
\label{g9}
   D_\mu(A) = \partial_\mu + A_\mu~,\quad
   \tilde F_\mu = {1\over 2} \epsilon_{\mu\nu k\gamma} 
   F_{k \lambda}~.
\ee
By minimizing the Yang-Mills action 
\be 
\label{g10}
  S_{YM} = {1\over 4k^2} \int \left( F(A)\right)^2
\ee
one finds the classical Yang-Mills equation of motion
\be 
\label{g11}     
   \left[ D_\mu(A), F_{\mu\nu}\right] =0~,
\ee
the solutions of which are the well-known instantons. 

On the other hand, in the field strength approach the fundamental
quantity is the tensor field $\chi_{\mu\nu}^a(x)$ and from this
the induced gauge field (\ref{g6}) is built up. By construction
this induced gauge field satisfies the equation of motion 
\be 
\label{g12}
    \left[ D_\mu(V), F_{\mu\nu}\right] =0   
\ee
but will in general not satisfy
the Bianchi identity. However, the classical equation of motion
obtained by minimizing the action (\ref{g5}) reads 
\be 
\label{g13}
   \chi = iF_{\mu\nu}(V)~,
\ee
which shows that the classical tensor fields are in
turn (up to a
constant) equivalent to the field strength constructed 
from the induced gauge field (\ref{g6}). This implies that the
classical tensor fields in fact satisfy the Bianchi identity
\be 
\label{g14}
     \left[ D_\mu(V), \tilde \chi_{\mu\nu}\right] =0~.
\ee     
Therefore we observe that in the field strength approach the
roles of the classical equation of motion and the Bianchi
identity are interchanged compared to the original Yang-Mills
theory. This is an essential feature of a dual
formulation. Moreover, as it is clear from equation (\ref{g3}),
in the field strength formulation the coupling constant is
inverted, compared to the original theory. This, together with
the above observation justifies calling the field strength
approach the dual formulation of Yang-Mills theory, although it
is not formulated in terms of a dual potential. This, however,
might be an advantage. 

Let us also mention, if one applies the field strength approach
to QED one obtains in 
fact the dual QED. Furthermore, for compact QED the field
strength approach yields the $\Zset$~ gauge theory. 

The semiclassical analysis of Yang-Mills theory can be performed
in the field strength approach in the same way as in the
standard formulation. In fact the stationary points of the field
strength action (\ref{g5}) are given by $iF_{\mu\nu} [A_\mu^{\rm
inst}]$ where  $A^{\rm inst}$ denotes the instanton gauge
potential and the corresponding induced gauge field (\ref{g6})
becomes the 
instanton field. Furthermore, calculation of the leading
quantum fluctuations~\cite{r9} yields the same result as in
the standard approach.

\section{Field strength approach to D=3 Yang-Mills theory in gauge
invariant variables}

In D=3 dimensions the tensor field can be expressed in terms of
a color vector 
\be \label{g15}
   \chi_{ij} = \epsilon_{ijh} \chi_k^a
\ee
which transforms homogeneously under gauge transformations. In
terms of the color vector $\chi_k^a$ the induced gauge potential
(\ref{g6})
becomes precisely  the representation of the gauge potential
introduced by Johnson et al. in their gauge invariant
formulation of Yang-Mills theory in the hamilton 
approach~\cite{r10}. We can use their result 
to formulate the field strength
approach to three-dimensional Yang-Mills theory in terms of
gauge invariant variables. For this purpose we introduce the
gauge invariant metric 
\be \label{g16}
   g_{ij} = \chi_i^a \chi_j^a~.
\ee
In the same way as in gravity we introduce the affine connection 
\be \label{g17}
   \Gamma_{jk}Π= {1\over 2} g^{im}
   \left(\p_j g_{mk} + \p_k g_{jm} - \p_m g_{jk} \right)~.
\ee
From this we construct the Riemann curvature 
\be \label{g18}
   R_{kij}^\ell = \p_i \Gamma_{jh}^\ell -\p_j \Gamma_{jk}^\ell
   + \Gamma_{jk}^m \Gamma_{im}^\ell - \Gamma_{ik}^m
   \Gamma_{jm}^\ell~.
\ee
Further, defining the Ricci curvature and the corresponding
Ricci scalar
\be \label{g19}
   R_{k\ell} = R_{ki\ell}Œ~, \quad R=R_k^k~,
\ee
and the Einstein curvature
\be \label{g20}
   G_{k\ell} = R_{k\ell} - {1\over 2} g_{k\ell} R~,
\ee
one can express the field strength of the induced gauge
potential by 
\be \label{g21}
   F_{ij}^a(V) = \epsilon_{ijk} B_k^a~,
   \quad B^{ai} = \sqrt g \chi_j^a G^{ij}
\ee
where $\sqrt g = det~\chi_i^a$. Using furthermore that in D=3:
$det~\hat\chi = -2(det~\chi_i^a)^3$, for vanishing external
sources $j=0$ the functional integral (\ref{g4}) can be entirely
expressed in terms of the gauge invariant metric (\ref{g16}).
One obtains 
\be \label{g22}
   Z[j=0] = \int \cD g_{ij} g^{-2} \exp \left[
   -{k^2\over 2} \int g_{ii} + {i\over 2} \int \sqrt g
   g_{ij} G^{ij} \right]~.
\ee
Here the kinetic term of the gauge invariant metric (the last
term in the exponent) coincides with the action of D=3 gravity.
On the classical level the correspondence between
three-dimensional Yang-Mills theory and gravity was also
observed in~\cite{r11}.

\section{Field strength apporach in the   Maxwell gauge}

The realization of the dual Meissner effect assumes the
existence of magnetic monopoles which can be most easily
identified by using 't Hooft's Abelian projection~\cite{r12}. 
This is based on the Cartan decomposition of the gauge group 
$G=H\otimes G/H$ where $H$ denotes the Cartan subgroup
(invariant torus). Accordingly the gauge potential can be
decomposed into a part $A_\mu^n$ living in the Cartan subalgebra
and a part $A_\mu^{ch}$ living in the coset $G/H$. Lattice
calculations~\cite{r13} indicate that there is a preferred
maximal abelian gauge 
\be \label{g23}
   \left[ \p_\mu +A_\mu^n~,~A_\mu^{ch} \right] =0~.
\ee
In this gauge the monopoles seem indeed  to be the relevant
infrared degrees of freedom. One therefore would like to have an
effective Abelian theory with magnetic monopoles present, where
however the charged field $A_\mu^{ch}$ is completely integrated
out. This, in fact, can be done in the field strength approach at
the expense of an Abelian tensor field. This is achieved by
linearizing not the complete field strength as it was done in
(\ref{g3}), but only the non-Abelian part of the field strength 
$\left[ A_\mu^{ch}~,~A_\nu^{ch}\right]$. For the gauge group
$G$ = SU(2) this commutator is within the Cartan algebra, and
accordingly, the tensor field $\chi_{\mu\nu}$ necessary to
linearize the square 
of this term (see eq. (\ref{g3}) lives also in the Cartan
subalgebra. After integrating out the charged gauge field
$A_\mu^{ch}$ one is left then with an effective theory in the
Abelian gauge field $A_\mu^n$ and an Abelian tensor field 
$\chi_{\mu\nu}$. The explicit form of this effective theory has
been presented in~\cite{r14}. Before extracting the
nonperturbative physics from this effective theory we have
calculated the one-loop Beta function~\cite{r15} and reproduced
the standard result. At present we are searching for nontrivial
tensor field configurations which would induce interactions
between the magnetic monopoles. 

\bi
\section*{Acknowledgements}

Discussions with K. Johnson and M. Quandt are gratefully
acknowledged. 

\no
This work was supported in parts by DFG Re 856/1-3 and funds
provided by the U.S., Department of Energy (D.O.E.) under
cooperative agreement \\ \# DE-FC02-94ER40818.

\bi
\no
{\em Note added:} After this work was completed, Ref.~16 was
brought to our attention, which also uses the variables
introduced in Ref.~10 to formulate the field strength
approach in terms of gauge invariant variables.

\end{document}